# Pushing for higher rates and efficiency in Satcom: the different perspectives within SatNExIV

Miguel Ángel Vázquez[1], Ana Pérez-Neira[1,2], Carlos Mosquera[3], Bhavanni Shankar[4], Pol Henarejos[1], Athanasios D. Panagopoulos[5], Giovanni Giambere[6], Vasilios Siris[7], George Polyzos[7], Nader Alagha[8].

Centre Tecnològic de les Telecomunicacions de Catalunya, Castelldefels, Barcelona, Spain[1].
Universitat Politècnica de Catalunya, Barcelona, Spain[2].
Universidad de Vigo, Vigo, Spain[3].
Interdisciplinary Centre for Security, Reliability and Trust, (SnT), University of Luxembourg, Luxembourg[4].
National Technical University of Athens, School of Electrical & Computer Engineering, Athens, Greece[5].
CNIT - University of Siena Department of Information Engineering Siena, Italy[6].
Mobile Multimedia Laboratory, Athens University of Economics and Business, Athens Greece[7].
European Space Agency (ESA-ESTEC), Noordwijk, Netherlands[8].

*Abstract*— SatNEx IV project aims at studying medium and long term directions of satellite telecommunication systems for any of the commercial or institutional applications that can be considered appealing by key players although still not mature enough for attracting industry or initiating dedicated ESA R&D activities. This paper summarizes the first year activities identified as very promising techniques for next generation satellite communication systems. Concretely, very high throughput satellite trunking, physical layer advances for full-duplex and multipolarization systems, network coding applications and multiple access schemes for information centric networking are briefly presented. For all the activities, we identify the scenarios under study so as the preliminary technical solutions to be further investigated.

## I. Introduction

Satellite NEtwork of Experts IV (SatNEx IV) is a network of excellence funded by the European Space Agency (ESA) with the objective of investigating future satellite telecommunications systems. The network aims at i) early identification, exploration and scientific assessment of promising new R&D avenues for satellite telecom networks for possible injection in ESA's R&D programs; ii) detection and preliminary assessment of promising terrestrial telecommunication technology spinning into space telecom applications; iii) enhance cooperation between the European/Canadian industry and research institutions on telecom satellite applied research subjects of common interest. In order to deal with the aforementioned objectives, SatNEx IV will yearly fund a set of 4 or 5 activities coined as Work Items (WI) that will investigate early satellite communications technologies and systems. These activities will be performed by certain members of the network with the supervision of ESA and with the support of certain satellite industries. Furthermore, in order to foster a prolific collaboration between the European satellite research community, dissemination activities will be held together with a seasonal school.

This paper describes the very first outcomes of the project so as the first year activities to be performed. During this first period, the objectives of the network of excellence have been to identify the most promising scenarios and techniques to be investigated. In addition, certain preliminary solutions have been provided.

The rest of the paper presents the four WI that are being investigated this year. In particular, Section II describes the study of next generation satellite trunking. Section III, identifies the physical layer advances for mobile and fixed communications. Section IV describes a survey on future network coding (NC) applications for satellite systems and Section V proposes to revisit the medium access schemes when considering information centric networking (ICN).

## II. Next Generation Satellite Trunking

### A. Objectives

The objective of this work item is the investigation and the study of the next generation of ultra-high capacity satellite constellations dedicated to satellite trunking. More particularly, in WI1, a consolidated system study will be considered taking into account propagation inputs in order to accurately dimension the next generation satellite trunking systems. It will be considered mostly NGSO solutions such as MEO satellites operating either in RF bands, i.e., Q/V bands, or in optical bands. Most specifically the reference system scenario will be a MEO constellation with 8 satellites operating at Ka band. For the evaluation of the next generation satellite trunking systems, time series synthesizers for the propagation phenomena for RF (Q/V bands) and optical systems are required.

To sum up, the main tasks to be addressed in WI are a) Satellite Constellation, b) Ground Network, c) Link Budgets, d) Physical Layer, e) Advanced diversity handover concepts and f) Antenna design.

## B. Definition of the scenarios

The operation frequencies and the allocated bandwidth of the trunking satellite systems will be investigated considering the ITU-R regulations for RF systems and the optical windows for the optical links. For RF systems, the study will be concentrated in the Q/V bands. Moreover, considering the space and ground segment of the system, MEO satellites will mainly be employed for this study and, therefore, the constellation size, the altitude and the inclination angles will be investigated in order to have quasi-global coverage, and a minimum latency (nominally, the round-trip is considered without considering outage due to propagation phenomena). The feasibility of the intersatellite links (ISL) in terms of acceptable latency based on the latitude of the user terminals will be investigated. Moreover, the coverage of the satellite systems and the latency will be studied in terms of the number of ground stations and the link budgets.

## C. Channel modelling

For the accurate evaluation of next generation satellite trunking systems, accurate total attenuation synthesizers are used for the MEO RF systems. The channel models are based on Stochastic Differential Equations (SDEs) VI. More particularly, time series of rain attenuation based on SDEs for single link are generated. There will be an extension for multiple ground stations (parallel slant paths) and multiple satellites (convergent links) using validated spatial correlation with multi-dimensional SDEs. As to the clouds attenuation, multi-dimensional SDEs will be used to generate the corresponding time series considering ITU-R. P.840 (Rayleigh approximation) and ITU-R. P. 1853 for RF Q/V bands links. Regarding the turbulence time series, SDEs driven by fractional Brownian motion are used [4]. For correlated total attenuation for water vapor and the cross-correlation between the attenuation factors, ITU-R. P. 1853 will be adopted. For the optical satellite links, firstly, the dimensioning of the system is based on the Cloud Free Line of Sight (CFLOS). The choice of the ground stations locations is derived through the desired availability of the system. The probability that a number of stations is under no clouds can be calculated through two different methods: either through data from ECMWF ERA-40 or ERA-Interim considering the total cloud coverage, or through the definition of the joint probability that the integrated Liquid water content (L) is above 0 mm. The spatial correlation coefficient proposed in [5] will be used. For the channel model of optical turbulence, SDEs will also be employed and plane, spherical and Gaussian beam waves will be considered [6], [7].

### III. ADVANCE SIGNAL PROCESSING TECHNIQUES

This WI aims at investigating the future advances in physical layer satellite communications systems. Precisely, multipolarization transmission in both fixed and mobile scenarios; and full duplex communications are going to be investigated.

## A. Mobile Interactive Satellite Communications

Increasing the bit rate of multimedia mobile satellite scenarios is becoming a challenge and requires new paradigms to fulfil the requirements of incoming technologies. Among all the possibilities, one of the most promising is the polarization state of the waveform to convey information. For instance, Polarization Multiplexing (PM) [8][9]; Orthogonal Polarization Time Block Codes (OPTBC) can be used to exploit the polarization diversity [10]; and as most recent works unveil, Polarized Modulation (PMod) can be used to increase the robustness in front of cross polarization discrimination (XPD) [11].

In parallel to the recent trends in signal processing, new standardization bodies are encompassing these ambitions with the current deployments. Particularly, the new standard Broadband Global Area Network (BGAN) aims to offer multimedia communications in hand held terminals in S/L bands, with very low latencies, and increase the network capacity [12].

This work item pursues the enhancement of sum rate techniques providing the following aspects: i) multiple polarization schemes: single, OPTBC, PM and PMod; ii) dynamic adaptive modulation and coding but also adaptive polarization scheme; iii) physical Layer Abstraction for system level simulations using the BGAN standard in realistic scenarios; iv) deployment of reliable time series dual polarization mobile satellite channel.

## B. Poly-polarization

Polarization has been used traditionally as an additional resource (colour) towards enhancing the throughput of the system. In a significant step forward, polarization can be seen to provide the MIMO paradigm for satellite communications since spatial MIMO is not attractive due to the nature of the channels [13]. Dual polar MIMO has been proposed for Mobile Broadcasting extending standard DVB-SH [14]. Several space-time techniques like Alamouti, Golden codes have been considered and significant gains over the single input single output channel have been demonstrated for measured channels in certain scenarios. However, such polarization based MIMO transmission from satellite to ground terminal provides significant gains when the channel is not completely Line of Sight. In particular, in case of fixed applications employing higher bands, Line of Sight dominates the other effects and use of polarization to enable MIMO seems less rewarding.

To cater to scenarios complementary to MIMO, a novel mechanism for exploiting polarization has been proposed [15], [16]. Termed Polypolarization Multiplexing (PPM), the transmission mechanism is shown to provide gains over the traditional orthogonal multiplexing system when employed on LoS channels. Gains to the tune of 0.5dB have been obtained for uncoded systems [16] and a system demonstrator has also been devised for the PPM [17], [18]. The current activity builds on the PPM designs and investigates the possibility of further optimization as well as impact of satellite channel on the new designs. Focussing on the forward link of a Fixed

Satellite Service (FSS), the key objectives explored in this work item include an exploration of the multidimensional constellation design paradigm to enhance spectral efficiency when using more than one polarization, establishing theoretical limits on the performance enhancements and undertaking system level performance analysis incorporating channel impairments such as cross-polarization, non-linearities (as and when deemed appropriate).

### C. Full-duplex

Full Duplex (FD) represents an attractive solution to improve the throughput of wireless communications. The term FD is historically used to refer to those systems that transmit and receive simultaneously. If transmission and reception take place in the same frequency band, then In-Band Full Duplex (IBFD) is a more precise term [20]. IBFD has the potential of doubling the spectral efficiency with respect to Half-Duplex (HD). In the case of Satellite Communications, current regulation is such that different frequency bands are allocated for uplink and downlink connections with a few exceptions, such as the Iridium mobile service in the L band. Time Division Duplexing (TDD) is used in the user link, with both transmission directions alternatively used in time.

The main challenge of IBFD is the attenuation of the Self-Interference (SI) caused by the transmission, so that the coupling does not spoil the reuse of the frequency band. Different methods have been devised to reduce the echoed signal power; passive techniques, some of them based on advance antenna design concepts, and active cancellation (both in the analog and digital waveform domains) can provide altogether the required isolation in some cases [21]. FD operation may also need to resort to cancellation countermeasures if transmission and reception bands are not fully decoupled, even though they are disjoint [22]. In satellite communications, the expected high imbalance between transmit and receive powers impose important challenges.

Some mechanisms have been proposed to reuse the same frequency band from two different sites that communicate through the satellite. These systems, known at the commercial level as PCMA (Paired Carrier Multiple Access) and Carrier-in-Carrier, can be considered as different implementations of Analog Network Coding (ANC), a kind of NC implemented at the physical layer [23], and must not be confused with FD systems despite their need to cancel the local replica of the transmitted signal at the waveform level [24], [25].

In this line of work the application of cancellation techniques is explored for some emerging satellite scenarios for which proper isolation between transmission and reception signals on the satellite cannot be guaranteed due to the proximity of the corresponding frequency bands. Different cancellation mechanisms are considered with the on-board satellite complexity as a major constraint. The successful attenuation of the local echo at an affordable cost could increase the flexibility of the allocation of frequency bands.

## IV. NETWORK CODING APPLICATIONS

### A. Motivation

NC has shown many potential advantages in protecting from packet erasures, in allowing an efficient delivery of contents to multiple users, and in permitting cooperative scenarios in the delivery of information [26]. The main idea with NC is to allow nodes in the network to perform coding operations at the packet level. There are different MC types, such as deterministic linear network coding, Random Linear Network Coding (RLNC), convolutional NC, etc.

The Internet Research Task Force (IRTF) has approved the Network Coding Research Group (NWCRG), considering several areas to investigate the applicability of NC to the networks [27], such as: architectural considerations (control plane, routing plane, transport layer, etc.), end-to-end versus hop-by-hop NC, inter-flow NC (at layer 3 or below) versus intra-flow (at layer 4 or above), application-layer NC, security issues and robustness to attacks, packet formats, proactive NC protection versus reactive ARQ- based mechanisms (i.e., trade-off between redundancy and retransmission delays).

### B. Technical Challenges

This activity will address satellite network applications of NC used at different OSI layers of the protocol stack, according to the following technical challenges.

**Network coding adopted at application layer**. This is an end-to-end case for NC use.

**Network coding applied at transport layer.** We will study advanced intra-flow NC techniques applied to multipath satellite scenarios (multi-homing), where end-user terminals can simultaneously dispose of multiple access technologies (heterogeneous network scenario). Moreover, we will also investigate reliable multicast protocols to include error recovery features coupled with ARQ schemes.

**Network coding applied at network layer**. In this case, network coding is applied to the communications between routers (hop-by-hop) with the possibility of recoding at intermediate nodes. We will investigate the NC interactions with different protocol types, such as: IPv4/IPv6 mobility proto- cols, GSE encapsulation, IP header compression (e.g., ROHC), etc.

### C. Scenarios

This WI surveys scenarios (and related OSI layers), where the NC adoption can allow achieving performance improvements. This would permit us to select suitable scenarios in relation to the above TCs to implement simulators for numerical assessments. Our main interest in this activity is on the applications concerning RLNC. A preliminary investigation of the literature has permitted to identify many NC applications Many of these studies do not refer to the satellite network scenario. In order to fill the research gap, in this activity we have identified three multipath NC scenarios to be studied by means of simulations as follows:

**Multicast transmission with satellite/terrestrial components.** Satellite transmissions can be impaired by many obstacles that are very common in the cities. To overcome this problem in vehicular IP multicast applications via satellite, the use of terrestrial gap-fillers or complementary wireless networks has been proposed in the DVB-SH standard. We will study a satellite that multicasts data to multiple terrestrial nodes and Road Side Units (RSUs); data packets are coded together by means of NC, applied before the transport level. RSUs, equipped with DVB-SH and 802.11p interfaces, cooperate in propagating the information received on the DVB-SH interface, re-transmitting it on the 802.11p interface.

**Satellite-based aeronautical applications.** We will study the implications of aircraft mobility when NC is applied to satellite mobile unicast and multicast for aeronautical communications, referring to the scenario of handover of multihomed aircraft mobile terminals. A simulation study will consider a dual frequency (L- and Ka- bands) aircraft terminal and investigations will be carried out for both forward and return links.

**Multipath TCP-based connections for satellite networks.** Study of RLNC performance in relation to Multi-Path TCP (MP-TCP) -based connections with simultaneous use of two (satellite) links. We consider that satellite terminals (mobile users, ON/OFF satellite channel) are able to receive traffic from two paths. A novel Path-Based Network Coding scheme for MP-TCP (PBNC-MP-TCP) will be investigated and compared with suitable benchmark techniques. With PBNC-MP-TCP, we adopt a sort of "coding diversity" where information packets are sent on a primary path and redundancy NC packets are sent on a secondary path so when bad channel conditions affect the primary path, we can take advantage of the other path and related redundancy packets. A possible approach is that an intermediate router supports Performance Enhancing Proxy (PEP) functionalities with subflow-level NC (shim layer) in order to implement PBNC-MP-TCP inside the network.

## V. IMPACT OF FUTURE ICN TRAFFIC ON MULTIPLE ACCESS SCHEMES

Research on ICN as a future Internet architecture has focused mainly on terrestrial networks. Among the few research activities on ICN and satellite networks, ESA's ARTES-1 φSAT has investigated the application of ICN architectures for the integration of satellite and terrestrial networks, and their corresponding advantages, disadvantages, and tradeoffs,. The introduction of ICN for satellite-terrestrial network integration has among others the following advantages: 1) Exploitation of ICN multipath mechanisms together with content-based routing and in-network caching allows different types of traffic to be routed along different paths that can include high delay satellite links and lower delay terrestrial links, thus permitting the separation of data and control traffic, as well as supporting differentiated QoS provisioning and flexible utilization of satellite capacity. 2) Management of the heterogeneity of physical layer characteristics across end-to-end paths spanning integrated satellite-terrestrial networks, through the separation of routing and forwarding that allows different forwarding mechanisms to be applied to different network segments based on their particular characteristics and requirements. 3) Improved security and privacy: ICN supports content-based security and, being receiver-driven, ICN offers by design significant advantages in terms of reduced spamming and protection against Denial-of-Service (DoS) attacks.

The overall goal of this work item is to investigate the interplay of ICN architectures and satellite multiple access schemes, in order to support M2M/IoT services. The specific objectives are: i) design of an end-to-end functional system architecture for an integrated satellite-terrestrial network based on ICN/PSI (Information-Centric Networking/Publish-Subscribe Internetworking) supporting M2M/IoT services, ii) investigation of the impact of M2M/IoT traffic on multiple access schemes, iii) test-bed validation of ICN/PSI architectures over enhanced satellite multiple access schemes, and iv) optimization of critical network functions of ICN/PSI architectures over enhanced satellite multiple access schemes.

The investigations will consider scenarios such as synchronous software upgrading of M2M/IoT nodes using GEO satellites, and the interconnection of massively deployed IoT sensors over LEO and over hierarchical LEO/MEO/GEO satellite architectures. The scenarios involve characteristics and requirements that can highlights the advantages of using ICN for the integration of satellite and terrestrial networks, which include wide-scale deployment and remote coverage, scalable one/many/any-to-one/many/any content dissemination and collection, dynamic network topology (in the case of LEO and MEO satellites), and content-based traffic management and service differentiation. The performance for the various scenarios will be assessed in terms of the reliability, scalability, and signalling overhead, identifying their corresponding tradeoffs and the impact of M2M/IoT traffic characteristics. Different categories of M2M/IoT traffic will be considered, such as software updates, periodic measurement reports from sensors, urgent alert data, and control messages.

The study will consider random access mechanisms based on slotted Aloha, such as Contention Resolution Diversity Slotted Aloha (CRDSA), fully scheduled (reservation-based) techniques, and spread spectrum techniques. An objective will be to identify the main features of each mechanism in relation to the M2M/IoT scenarios considered. For example, random access schemes are better suited for bursty traffic, deterministic access schemes are better suited for periodic reporting, while hybrid schemes are better suited for randomly generated software upgrades.

Regarding ICN architectures, the study will investigate the signalling overhead, in terms of number and rate of control messages, for each of the three main ICN functions: name resolution, route (or topology) management, and forwarding. This investigation will identify the functions that present room for optimization. Moreover, we will propose signalling aggregation/grouping schemes that exploit ICN's key features, which include content-awareness, multipath/multisource transport, and receiver-driven (pull-based) data transfer, along with the inherent broadcasting/multicasting capabilities of satellite networks. The aggregation schemes will be evaluated for the different M2M/IoT scenarios and traffic characteristics.

The validation experiments will be conducted in an integrated test-bed that consists of a prototype ICN implementation (FP7 PURSUIT's Blackadder implementation of the PSI architecture) and an open source satellite emulation module (OpenSAND), illustrated in Figure 1. The test-bed provides an extendable evaluation platform and includes modules implementing content-based traffic engineering (e.g. route selection), multipath and multisource transport, and content-aware security functionality.

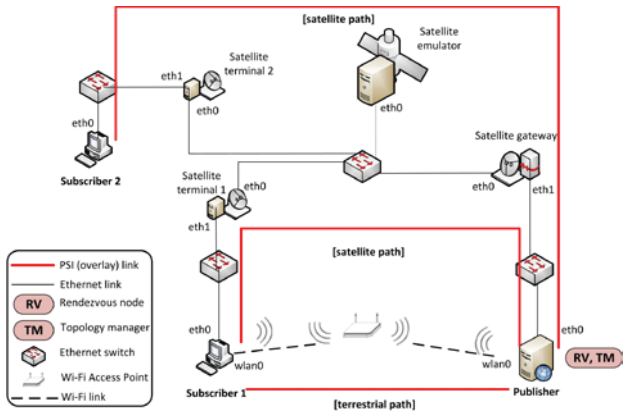

Fig. 1.  ICN Testbed